\documentclass{article}
\usepackage{psfig}

\newcommand{\be}{\begin{equation}}
\newcommand{\ee}{\end{equation}}
\newcommand{\bea}{\begin{eqnarray}}
\newcommand{\eea}{\end{eqnarray}}
\newcommand{\der}[2]{\frac{\partial #1}{\partial #2}}
\newcommand{\drf}[2]{\frac{\delta #1}{\delta #2}}
\newcommand{\dep}[3]{\frac{\partial^2 #1}{{\partial #2}{\partial #3}}}

\newcommand{\ot}{{\cal O}}
\newcommand{\veps}{\tau}

\title{$O(N)$ models within the local potential approximation}
\author{\\\\Jordi Comellas\thanks{\tt comellas@sophia.ecm.ub.es}
  \ \ \ \ Alex Travesset\thanks{\tt alex@greta.ecm.ub.es}\\\\
  \sl Departament d'Estructura i Constituents de la Mat\`eria\\
  \sl Facultat de F\'\i sica,\ Universitat de Barcelona\\
  \sl Diagonal, 647,\ 08028 Barcelona,\ Catalonia,\ Spain\\\\\\}
\date{}
\begin{document}
  \begin{titlepage}
    \maketitle
    \thispagestyle{empty}
    \begin{abstract}
      Using Wegner-Houghton equation, within the Local Potential Approximation,
      we study critical properties of $O(N)$ vector models.  Fixed Points,
      together with their critical exponents and eigenoperators, are
      obtained for a large set of values of $N$, including $N=0$ and
      $N\rightarrow\infty$.  Polchinski equation is also treated.
      The peculiarities of the large $N$ limit,
      where a line of Fixed Points at $d=2+2/n$ is present,
      are studied in detail.  A derivation of the equation is presented
      together with its projection to zero modes.
    \end{abstract}
    \vfill
    UB-ECM-PF 96/21\newline
  \end{titlepage}
  \section{Introduction}\label{SECT__Introduction}

    In this article we study $O(N)$ linear sigma models near
    criticality, within the framework of the Exact Renormalization
    Group (ERG, hereafter), mainly in its simpler nonperturbative
    approximation, the Local Potential Approximation (LPA). We 
    concentrate mostly in the Wegner-Houghton equation \cite{WH},
    although other equations are also considered 
    \cite{Polchinski,tim:ijmp}.

    Recent results \cite{SOK} have raised interesting questions about the 
    behaviour of the RG near first-order phase transitions.  They are in
    some conflict with exact results of Wegner-Houghton equation 
    at $N\rightarrow \infty$ \cite{HAS}.
    A systematic study of this subject at finite $N$ is still lacking. 
    Although our interest points in that direction, many
    aspects of second order phase transitions 
    had to be previously worked out.
    This paper is meant to fill this gap, as well as review the
    derivation and projection
    of the equation with a modern language, paying special attention to some
    aspects not fully treated
    in the literature, which we hope will be useful for a subsequent
    study of $O(N)$ models at discontinuous phase transitions.
    Nevertheless, as this paper is focused in continuous transitions,
    we leave to the conclusions further comments on first-order ones.

    ERG methods are exact formulations of the RG in differential form
    (see Ref.\ \cite{WILS}), that is,
    the evolution of the renormalized
    action along the RG flow is studied in terms of differential
    equations.
    There are many such equations, 
    being one of the oldest that of
    Wegner and Houghton \cite{WH},
    which corresponds to a decimation in momentum space,
    i.\ e., a lowering of a sharp momentum cutoff.
    Unfortunately, those equations are very involved and 
    exact solutions are only available for very simple models.
    Nevertheless, several feasible nonperturbative approximations are
    possible, being
    the most promising an expansion in powers of momenta.
    The LPA \cite{HH} is just
    the first term of such an expansion \cite{tim:LPA}.

    The organization of the paper is as follows.  In Section  
    \ref{SECT__Deriv} Wegner-Houghton equation is derived in some
    detail.
    The LPA is discussed in Section \ref{SECT__Projection} and extensively
    worked out in Section \ref{SECT__Results}, where we compute
    critical indices of $O(N)$ models at finite $N$ as well as
    its eigenoperators.  It is also shown that
    irrelevant eigenvalues, which are related to the
    anomalous dimensions of higher-dimensional
    composite operators,  may be obtained with ease.
    The comparison with our results from Polchinski equation
    as well as other determinations is left to Section \ref{SECT__Comparison},
    while the special, exactly solvable, $N \rightarrow \infty$ case
    is studied in Section \ref{SECT__LargeN}, where some striking  
    results are obtained. Section \ref{SECT__Conclusions}
    is devoted to the conclusions.
    Finally, some peripheral subjects are treated
    in two Appendices.

   \section{Derivation}\label{SECT__Deriv}

    In this section a careful derivation of Wegner-Houghton
    equation is presented, trying to be as self-contained as
    possible.
    Nevertheless, for the sake of simplicity, we concentrate on the $N=1$ case.
    Once this case is mastered, its generalization for arbitrary $N$ is
    straightforward.

    We assume that our action is regulated with a $t$-dependent sharp
    cutoff in momentum space
    $\Lambda_t=e^{-t}\Lambda_0$, with all modes with $|q|>\Lambda_t$
    already eliminated, where $t$
    parametrizes the RG flow and $\Lambda_0$ is a fixed scale.
    A RG transformation consists of two steps, a blocking 
    or elimination of short-distance degrees of freedom, and a change of
    length scale.  In this way one goes from an action
    cutoff at $\Lambda_t$ to one at $\Lambda_{t+\veps}$, in physical units.
    From now on, we work with dimensionless quantities, the dimensions being
    given, if needed, by $\Lambda_t$.
    We derive Wegner-Houghton equation first considering the blocking,
    and afterwards the rescaling.

    Let us parameterize our action as
    \bea\label{full_action}
     S[\phi] &=&
             \sum_{m=2,4,\ldots} \int_{q_1}\cdots\int_{q_m}
     v_{m}(q_1,\ldots, q_m)\phi_{q_1}\cdots\phi_{q_m} \delta_{q_1+\ldots+q_m}
	\nonumber
	\\
      &\equiv&\frac{1}{2}\int_qv_2(q)\phi_q\phi_{-q}+\tilde S[\phi],
    \eea	
    where $\int_q\equiv\int_{q\le 1}\frac{d^dq}{(2\pi)^d}$ and
    $\delta_q\equiv(2\pi)^d\delta(q)$.
    We have assumed
    \(
      Z_2
    \)
    symmetry.
    The field is split as
    \be\label{split}
    \phi_q=\phi^{(0)}_q+\phi^{(1)}_q,
    \ee
    where $\phi^{(0)}_q$ are the modes with $|q|\le e^{-\veps}$ and 
    $\phi^{(1)}_q$ the ones with $e^{-\veps}<|q| \le 1$.
    The blocked action is then defined through
    \be\label{blocked}
    e^{-S_{\veps}[\phi^{(0)}]}\equiv \int {\cal D}\phi^{(1)} e^{-S[\phi]}.
    \ee

    As we are seeking a differential equation for the renormalized action,
    we need to consider just an infinitesimal blocking, hence we
    discard terms of order
    \(
      \veps^2
    \)
    or higher.
    For that purpose, it is convenient to
    rewrite the action, Eq.\ \ref{full_action},
    as
    \be\label{stil_exp_1}
	S[\phi]=S[\phi^{(0)}]+
	  \frac{1}{2}\int'_q \phi_q^{(1)}v_2(q)\phi_{-q}^{(1)}
        +\sum_{n=1}^{\infty}\frac{1}{n!} \int'_{q_1}\cdots\int'_{q_n}
         \phi_{q_1}^{(1)}
        \ldots\phi_{q_n}^{(1)} \tilde{S}^{(n)}_{q_1,\ldots,q_n},
    \ee
    with
    \be\label{def_sder}
    \tilde{S}^{(n)}_{q_1,\ldots,q_n}\equiv\left.
    \frac{\delta^n \tilde{S}}{\delta \phi_{q_1}
    \ldots \delta \phi_{q_n}}\right|_{\phi^{(1)}=0}
    \ee
    and primes in integrals meaning momenta restricted
    to the shell
    \(
      e^{-\veps}<|q|\le1
    \).

    Written in the above form, the action is suited for treating the
    \(
      \phi^{(1)}
    \)-integral
    in a diagrammatic
    expansion.
    Its Feynman rules are:
    \begin{enumerate}
      \item Any $n$-legs vertex contributes with
        $\int'_{q_1}\cdots\int'_{q_n}\tilde S^{(n)}_{q_1,\ldots,q_n}$,
        where every leg is labelled by a momentum $q_i$.
      \item Any propagator between a leg labelled by $q_1$ and another one
        labelled by $q_2$ is $\delta_{q_1+q_2}v_2^{-1}(q_1)$.
      \item All $q_i$-integrals are understood to affect the whole
        diagram and not only the contribution of its vertex.  That is,
        every delta function always simplifies an integral.
      \item Add the usual symmetry factors.
    \end{enumerate}
    
    Before going on we state two technical results.

    {\em Lemma 1.}  For any function
    \(
      f(q;P)
    \),
    analytic around
    \(
      |q|,|P|=1
    \),
    it is
    \be\label{master1}
      \int[dP]\int'_qf(q;P)\delta_{q+P}
      =\ot(\veps),
    \ee
    where $P$ stands for a finite number of momenta $p_i$, with
    $|p_i|\le1$, and $\int[dP]$ for integrals over all $p_i$'s.

    {\em Proof.}  Just integrate first over all $P$
    integrals.  QED

    {\em Lemma 2.}  For any function $f(q_1,q_2;P)$, analytic for any
    value of $P$ and
    at $|q_1|,|q_2|=1$, it is
    \be\label{master2}
      \int'_{q_1,q_2}\delta_{q_1+q_2+P}f(q_1,q_2;P)=\ot(\veps),
    \ee
    provided that $P=0+\ot(\veps)$.

    {\em Proof.}
    In spherical coordinates, integrating with respect to $|q_2|$,
    \bea\label{integral}
      \lefteqn{\int'_{q_1}\int'_{q_2}\delta_{q_1+q_2+P}f(q_1,q_2;P)}
      \\\nonumber&&
      =\int_{e^{-\veps}<|q_1|\le1}\frac{d|q_1|}{(2\pi)^d}|q_1|^{d-1}
        \int d\Omega_1\int d\Omega_2\,\delta(\Omega)
        f(|q_1|\hat q_1,|q_1+P|\hat q_2;P),
    \eea
    with $|q_1+P|$ being in the shell and $\delta(\Omega)$ standing
    for the delta functions that constrain angular variables.
    The integral over
    \(
      |q_1|
    \)
    will give a contribution of
    \(
      \ot(\veps)
    \).
    To convince ourselves that the rest of the integral is
    \(
      \ot(\veps^0)
    \)
    for
    \(
      P=0
    \),
    let us take
    \(
      \theta_i
    \)
    to be the angle between $P$ and $q_i$.
    In this case, the angular contribution of Eq.\ \ref{integral}
    is
    \bea
      \lefteqn{\int d\Omega_1\int d\Omega_2\,\delta(\Omega)=
        \int d\Phi_1\int d\Phi_2\int d\theta_1\int d\theta_2J
        \delta(\Phi_1-\Phi_2)}\nonumber\\
        &&\times\delta(|q_1|\sin\theta_1-|q_1+P|\sin\theta_2)
        \delta(P+|q_1|\cos\theta_1+|q_1+P|\cos\theta_2)
    \eea
    where
    \(
      \Phi_i
    \)
    is a short-hand for all the remaining angular variables and $J$ is
    the appropriate Jacobian.  From the last two delta functions,
    recalling that
    \(
      1-\veps<|q_i|\le1
    \),
    \be
      \theta_1=\pi-\theta_2+\ot(\veps),\ \ \
      P=0+\ot(\veps).
    \ee
    QED

    Let us now compute the leading contribution, ${\cal O}(\veps^D)$, of 
    a tree diagram assuming that each 
    n-vertex is ${\cal O}(\veps^{[n/2]})$,
    with $[z]$ being the lowest integer greater or equal than $z$,
    propagators being
    ${\cal O}(1/\veps)$ because of the delta function.  Then,
    \be\label{for}
      D=\sum_nn(I_{2n}+I_{2n-1})-P=\sum_nn(I_{2n}+I_{2n-1})-\frac{1}{2}
        \sum_nnI_n=\frac{1}{2}\sum_nI_{2n-1},
    \ee
    where $I_n$ is the number of $n$-vertices and $P$
    the number of propagators.
    Let us now put this result in a rigorous basis.

    {\em Lemma 3.}  The leading contribution in $\veps$
    of a tree-level diagram,
    $\ot(\veps^D)$ is $D=\frac{1}{2}\sum_nI_{2n-1}$.

\begin{figure}[t]
      \centerline{\psfig{file=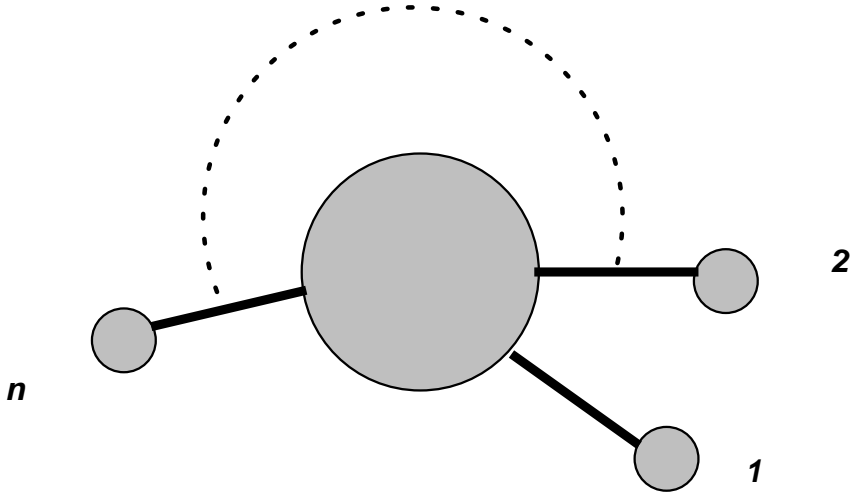,width=\textwidth}}
      \caption{A general diagram consisting on $n$ 1-vertices and
        1 $n$-vertex.}
	\label{fig__delta1}
    \end{figure}

    {\em Proof.}
    Consider first a connected tree diagram consisting of 
    one $n$-vertex and $n$ 1-vertices, as in Fig.\ \ref{fig__delta1}. 
    Its value is
    \be 
    \int'_{q_1}\cdots\int'_{q_n}{\tilde S}^{(n)}_{q_1,\ldots,q_n}
    \prod_{i=1}^{n}\frac{{\tilde S}'_{q_i}}{v(q_i)}.
    \ee
    We perform the $q_i$-integrals in pairs, using {\em Lemma 2\/} to extract
    the leading dependences in $\veps$.  At the end, we obtain,
    in agreement with {\em Lemma 3}, that it is
    $\ot(\veps^{n/2})$ for $n$ even or else $\ot(\veps^{(n+1)/2})$ for $n$ odd,
    with the final aid of {\em Lemma 1}.  (Recall the delta function
    hidden in Eq.\ \ref{def_sder}).

    Furthermore, let us assume that {\em Lemma 3\/} is true for all diagrams
    with $K$ vertices other than 1-leg ones.
    Let us now prove that it is also true for diagrams with $K+1$ such
    vertices.  To this end, note that a
    generic tree diagram ${\cal G}$ with $K+1$ more-than-one-leg vertices
    contains at least one $n$-vertex (for some $n\not=1$) with all 
    but one leg connected to 1-vertices.\footnote{
    To pick up one, just eliminate all 1-vertices, and
    take one 1-vertex of the remaining diagram (which must have at least
    one since it is again a tree).}

\begin{figure}[t]
      \centerline{\psfig{file=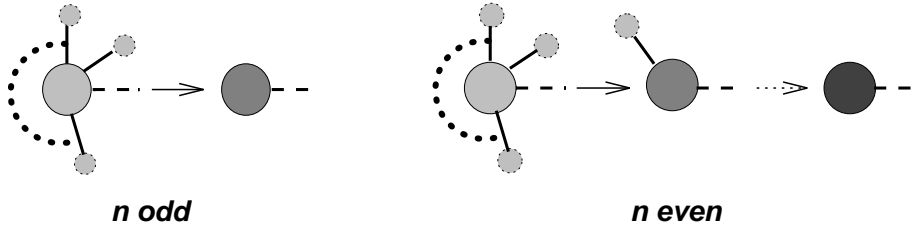,width=\textwidth}}
      \caption{A vertex with $n$ integrated legs, as explained in the proof
        of {\em Lemma 3}.  we have to distinguish between the $n$ odd and the
        $n$ even case.}
	\label{fig__odd_even}
    \end{figure}

    If the selected vertex is a n-vertex, with $n$ odd,
    perform
    all integrals associated with its legs but the one that connects it to the 
    rest of the diagram.
    Performing the integrals in pairs, and using {\em Lemma 2}, it
    brings about a contribution $\ot(\veps^{(n-1)/2})$.
    It remains a diagram $\cal G'$ with
    $K$ more-than-one-leg vertices,
    where the former selected vertex enters
    now as an additional 1-vertex (with
    $S'_{q}$ replaced by some other function).  See Fig.\ \ref{fig__odd_even}.

    If $n$ is even, then integrate over all its legs connected to
    1-vertices but one.  This is, using {\em Lemma 2\/} again,
    $\ot(\veps^{(n-2)/2})$.
    There is left over a 2-vertex connected
    both to the remaining sub-diagram and to a 1-vertex.  It can
    easily be seen that it is of the same order in $\veps$ as one with the
    1-vertex connected directly to the sub-diagram, without the 2-vertex.
    See again Fig.\ \ref{fig__odd_even}.

    Therefore,
    \be\label{ggprime}
    D({\cal G})=D({\cal G'})+\left[\frac{n-2}{2}\right]
    \ee
    and, by the induction hypothesis,
    \be\label{induction}
    D({\cal G'})=\frac{1}{2}\sum_{i}I_{2i-1}^{\cal G'}=
    \left\{\begin{array}{lc}
    \frac{1}{2}I_1^{\cal G'}+\frac{1}{2}\displaystyle{\sum_{i>1}}
      I_{2i-1}^{\cal G}
    & n \mbox{ even} \\
    \frac{1}{2}I_1^{\cal G'}+\frac{1}{2}\displaystyle{
      \sum_{
      i>1,i \not=\frac{n+1}{2}}}
    I_{2i-1}^{\cal G}+\frac{1}{2}I_n^{\cal G'} 
    & n \mbox{ odd} \end{array} \right.
    \ee
    Taking into account that $I^{\cal G'}_1=I^{\cal G}_1-(n-1)+1$ and
    $I^{\cal G'}_n=I^{\cal G}_n-1$, a proof of {\em Lemma 3\/} follows.

    QED

    Note that every tree diagram must necessarily be\footnote{
    Recall that every 
    diagram must contain at least one propagator.}
    \(
      I_1\ge2
    \),
    and, from {\em Lemma 3}, it must be
    \(
      I_1=2 
    \)
    at leading order in
    \(
      \veps
    \).
    Furthermore, in such case, any inclusion of a four-legs or higher
    vertex would imply at least one loop, and thus the
    only vertices allowed, apart from
    one-leg ones, are an arbitrary number of two-legs ones.  Even more, these
    additional vertices, in order to be
    \(
      {\cal O}(\veps)
    \),
    are ({\em Lemma 2})
    \be\label{constra}
    \int'_{q}\tilde S''_{q,-q}.
    \ee

    The argument that leads to the proof of
    {\em Lemma 3\/} has to be slightly modified
    for diagrams with $L$ loops.
    Any 1PI part contains one
    $q$-integral per loop, with the only restriction that
    of being within the shell. After performing those integrals, we end
    up with a contribution $\ot(\veps^L)$ times an effective tree diagram.
    Therefore, the correct formula is to
    add the number of loops to the result of {\em Lemma 3}, but
    with Eq.\ \ref{for}
    computed considering any 1PI part as a tree-level vertex.

    We conclude then that there cannot be diagrams with more than one loop and
    that one-loop diagrams must have zero external legs.  Furthermore,
    it happens again that diagrams with more than two legs generate further
    loops and that two-legs ones must have the form of Eq.\ \ref{constra}.

    All those considerations show that, at leading order in $\veps$,
    Eq.\ \ref{stil_exp_1} is equivalent to
    \be\label{stil_exp_2}
	S[\phi]=S[\phi^{(0)}]+
        \int'_{q} \phi_{q}^{(1)} S'_{q}
        +\frac{1}{2} \int'_{q} \phi_{q}^{(1)}
        \phi_{-q}^{(1)} S''_{q,-q}.
    \ee
    The path integral, Eq.\ \ref{blocked}, is now Gaussian with the result
    \be\label{integ_gaus}
       S_{\veps}[\phi^{(0)}]=
       S[\phi^{(0)}]-\frac{1}{2}\int'_{q}
       S'_{q}S'_{-q}\left(S''_{q,-q}\right)^{-1}+\frac{1}{2}\int'_{q}
       \ln\left(S''_{q,-q}\right)+\mbox{const},
    \ee
    where we assume that $S''_{q,-q}$ is positive definite and ``const''
    means a possible term independent of $\phi^{(0)}$.

     The other step in any RG transformation
     is a change of scale in all linear
     dimensions. This amounts to a standard dilatation transformation,
    \be\label{dilatation}
      k\equiv e^{\veps}q,\ \ \
      \phi_k\equiv e^{-\veps\frac{d+2-\eta}{2}}\phi^{(0)}_q,
    \ee
    which gives,
    \bea\label{rescal}
      S[\phi^{(0)}]&=&
        \sum_m\int_{k_1,\ldots,k_m}e^{-\veps md}v_m(e^{-\veps}k_1,\ldots,
        e^{-\veps}k_m)
        \nonumber
        \\
        &&
        \mbox{}\times
        \phi_{k_1}\ldots\phi_{k_m}e^{\veps m\frac{d+2-\eta}{2}}
        \delta_{k_1+\ldots+k_m}e^{\veps d}\\\nonumber
      &=&S[\phi]+\veps \left(
      dS-\frac{d-2+\eta}{2}\int_k\phi_k\frac{\delta S}{\delta\phi_k}
        -\int_k\phi_kk\cdot\frac{\partial'}{\partial k}
          \frac{\delta S}{\delta\phi_k} \right)+\ot(\veps^2),
    \eea
    where the prime in the last partial derivative merely indicates that it
    does not affect delta functions.

    Therefore, Wegner-Houghton equation is
    \bea\label{final_eq}
      \lefteqn{\dot S\equiv\der{S}{t}
      = \lim_{\veps \rightarrow 0} \frac{S_{\veps}[\phi]-S[\phi]}{\veps}}
        \nonumber\\
      &&=\lim_{\veps \rightarrow 0}
        \frac{1}{2\veps}\left[\int'_{k}\ln(
           S_{-k,k}'') 
          -\int'_{k} S_{-k}' S_{k}'
            \left(S_{-k,k}''\right)^{-1}
	\right]\\\nonumber
        &&\mbox{}+dS
          +\frac{2-d-\eta}{2}\int_k\phi_k\frac{\delta S}{\delta\phi_k}
          -\int_k\phi_k k\cdot\frac{\partial'}{\partial k}
            \frac{\delta S}{\delta\phi_k}+\mbox{const}.
    \eea
   
    To deal with the $N$ arbitrary case, 
    we define the partition function as
    \be\label{norm_N}
      Z[J^i]=\int{\cal D}\phi e^{-NS[\phi^i/\sqrt{N}]+\int J^i\phi^i},
    \ee
    so that couplings are all finite at
    \(
      N\rightarrow\infty
    \).
    The derivation is completely analogous and the ERG equation reads
    \bea\label{final_eqN}
      \dot S
      &=&\lim_{\veps \rightarrow 0}
        \frac{1}{2\veps}\left[\frac{1}{N}\int'_{k}\mbox{tr}\ln(
           S_{-k,k}^{j,i}) 
          -\int'_{k} S_{-k}^j S_{k}^i
            \left(S_{-k,k}^{j,i}\right)^{-1}
	\right]\\\nonumber
        &&\mbox{}+dS
          +\frac{2-d-\eta}{2}\int_k\phi_k^i\frac{\delta S}{\delta\phi_k^i}
          -\int_k\phi_k^ik\cdot\frac{\partial'}{\partial k}
            \frac{\delta S}{\delta\phi_k^i}
	  +\mbox{const},
    \eea
    where the trace is over flavour indices and 
    \be
    S^{i_1,\ldots,i_n}_{q_1,\ldots,q_n}\equiv \left. 
    \frac{\delta^n S}{\delta \phi_{q_1}^{i_1}
    \ldots \delta \phi_{q_n}^{i_n}}\right|_{\phi^{(1)}=0}.
    \ee

  \section{Projection}\label{SECT__Projection}

    The Local Potential Approximation (LPA) amounts to projecting the full
    action into a fixed
    kinetic term plus a general potential involving only the
    zero modes,
    \begin{equation}\label{action}
      S=\frac{1}{2}\int_k
	k^2\phi_k\phi_{-k}+V(y)\delta_0,
    \end{equation}
    where
    \(
      y\equiv\phi^i_0\phi^i_0
    \).
    The LPA is, thus, just the first term in
    an expansion in powers of momenta
    \cite{tim:LPA}.

    To project over constant fields, we define the operator
    \(
      {\cal P}(x^i)
    \)
    \cite{HH}
    which, acting on an arbitrary functional $G$, is
    \begin{equation}\label{proj_oper}
      {\cal P}(x^i)G=
	\left.\exp\left(x^i\frac{\delta}{\delta\phi^i_0}\right)G
	\right|_{\phi_k=0}.
    \end{equation}
    The projection of the third term in the second line of
    Eq.\ \ref{final_eqN}
    gives zero because of the explicit factor of $k$ there.  The first two
    terms in that line are simply
    \begin{equation}\label{proj_resc}
      dV(y)\delta_0
	+\frac{2-d-\eta}{2}x^i\frac{\partial V(y)}{\partial x^i}\delta_0=
      [dV(y)+(2-d-\eta)yV'(y)]\delta_0,
    \end{equation}
    with
    \(
      x^ix^i\equiv y
    \).
    The second term of the right hand side of Eq.\ \ref{final_eqN}
    can be seen to cancel
    whereas the first one is
    \be\label{eq_ant}
    \lim_{\veps\rightarrow0}
    \frac{1}{2N\veps}\int_k'\mbox{tr}\ln\left({\cal P}(x) S_{k,-k}^{i,j}
    \right).
    \ee
    The LPA amounts to
    \be\label{assumption}
    {\cal P}(x) S_{k,-k}^{i,j}\longrightarrow
    \left(\delta^{ij}k^2+\dep{V}{x^i}{x^j}\right)\delta_0.
    \ee
    Eq.\ \ref{eq_ant} becomes, thus,
    \begin{eqnarray}\label{proj_kad}
      \lefteqn{\frac{A_d}{2}\mbox{tr}\ln \left(\delta^{ij}+
	    \frac{\partial^2V(y)}{\partial x^i\partial x^i}\right)
	    +\mbox{const}} \nonumber \\
      &&=\frac{A_d}{2}\left[\mbox{tr}\ln(1+u)+
	  \mbox{tr}\ln\left(\delta^{ij}+2x^ix^j\frac{u'}{1+u}\right)\right]
	  +\mbox{const} \nonumber \\
      &&=\frac{A_d}{2}\left[(N-1)\ln(1+u)+\ln(1+u+2yu')\right]+\mbox{const},
    \end{eqnarray} with \( u(y)\equiv2V'(y) \),
    \begin{equation}\label{A_d}
      A_d\equiv\frac{2\pi^{d/2}/\Gamma(d/2)}{(2\pi)^d}
    \end{equation}
    and
    \(
      \Gamma(z)
    \)
    the Euler Gamma function.

    The equation is further simplified if we perform one more
    derivative with respect to $y$,
    \begin{equation}\label{proj_eq_del}
      \delta_0\dot u=\frac{A_d}{N}\left[\frac{3u'+2yu''}{1+u+2yu'}
	+(N-1)\frac{u'}{1+u}\right]
      +\delta_0[(2-\eta)u+(2-d-\eta)yu'].
    \end{equation}

    If, instead, the projection is
    over the kinetic term and its
    normalization kept fixed under the RG flow,
    \begin{equation}\label{ren_kin}
      0=-p^2+\frac{1}{2}dp^2+\frac{1}{2}(2-d-\eta)p^2,
    \end{equation}
    it follows
    \(
      \eta=0
    \).

    Moreover,
    the delta function in Eq.\ \ref{proj_eq_del}
    just reflects the difficulties of selecting one mode
    out of a continuum set.  This problem is not present at finite volume,
    where the zero mode is neatly separated.  In such a case the delta function
    $\delta_0$ is just a coefficient proportional
    to the inverse number of modes.  One can then rescale
    $y\rightarrow y/\delta_0$ and Eq.\ \ref{proj_eq_del} becomes
    \begin{equation}\label{proj_eq}
      \dot u=\frac{A_d}{N}\left[\frac{3u'+2yu''}{1+u+2yu'}
	+(N-1)\frac{u'}{1+u}\right]
      +2u+(2-d)yu'.
    \end{equation}
    Note that, along the same lines, one may absorb $A_d$.

   On projecting, we made just one approximation, that of
   Eq.\ \ref{assumption}. Before closing the section, we briefly
   analyze how, in the $N\rightarrow\infty$ limit,
   this approximation becomes exact.
   Eq.\ \ref{assumption} amounts to,
   \bea
    v_2(k) &\rightarrow& \frac{1}{2}k^2+v_2(0)
    \nonumber\\
    v_m(k,-k,0,\ldots,0) &\rightarrow& v_m(0,\ldots,0) \ \ \ m >2.
   \eea
   Hence, it becomes exact when
   couplings do not depend on $k$ (except for $v_2$ that has a $k^2$
   term). If one expands Eq.\ \ref{final_eqN} in powers of momenta,
   one realizes that an hypothesis of such a kind does not hold for $N$ finite,
   but there are some chances at $N=\infty$.

   To proceed, it is necessary to rewrite Eq.\ \ref{final_eqN}
   introducing $O(N)$ invariants,
   \be
   \Phi_{k_1,k_2}=\phi_{k_1}^i\phi_{k_2}^i.
   \ee
   Then, Eq.\ \ref{final_eqN} at $N=\infty$ becomes closed for the invariants
   of the form $\Phi_{k_1,k_2}=\delta_{k_1+k_2}\Phi_{k_1}$,
   \be\label{WH_mod}
    \dot{\hat S}=\lim_{\veps\rightarrow0}
    \frac{1}{2\veps} \int_{k}'\ln(2\drf{\hat S}{\Phi_k})
    +d\hat S+(2-d)\int_k\Phi_k\drf{\hat S}{\Phi_k}
    -\int_k \Phi_kk\cdot\der{'}{k}\drf{\hat S}{\Phi_k}+\mbox{const},
   \ee
   with $\hat S$ is the piece of $S$ that contains only the invariants
   $\Phi_k$ (called diagonal piece in Ref.\ \cite{WH}),
   and we set $\eta=0$ which is a general result on the 
   $N \rightarrow \infty$ limit.

    An inspection of how couplings of the diagonal
    piece at $k=0$ are affected by those
    with increasing powers of $k$ suggest that, if all couplings at
    a certain power of $k$ are set to zero, all the rest but the
    ones at $k=0$ can be simultaneously
    set to zero, so it should not come as a
    surprise that an ansatz like
   \be
   S[\phi]=\frac{1}{2}\int_k k^2\Phi_k+V(\varphi^2),
   \ee
   with $\varphi^2=\int_k \Phi_k$, holds.

   Indeed, Eq.\ \ref{WH_mod} leads to a closed equation for $V(\varphi^2)$,
   \be\label{exact}
   \dot V=A_d\ln\left(1+V'(\varphi^2)\right)+dV(\varphi^2)+(2-d)\varphi^2
     V'(\varphi^2)+\mbox{const},
   \ee
   which turns out to be
   the same as the large $N$ limit of Eq.\ \ref{proj_eq}. This
   is the result we were after. The LPA in the large $N$ limit is not only 
   solvable as we will see in Section \ref{SECT__LargeN},
   but also exact. If our initial
   potential has couplings that do not depend on $k$, the exact evolution 
   for those is governed by Eq.\ \ref{exact}.

  \section{Results}\label{SECT__Results}
    Fixed Points (FPs)
    \(
      u^*(y)
    \)
    of Eq.~\ref{proj_eq}
    are the analytical solutions of
    \begin{equation}\label{FP_eq}
      0=\frac{A_d}{N}\left[\frac{3u'+2yu''}{1+u+2yu'}
	    +(N-1)\frac{u'}{1+u}\right]
	+2u+(2-d)yu',
    \end{equation}
    with
    \(
      A_d
    \)
    some $d$-dependent constant (Eq.\ \ref{A_d})
    which
    equals
    \(
      (2\pi^2)^{-1}
    \)
    in
    \(
      d=3
    \).
    All universal quantities are independent of its precise value.
    Note that this is an ordinary differential equation of second order and,
    thus, two initial conditions are required.
    Nevertheless,
    at
    \(
      y=0
    \)
    Eq.\ \ref{FP_eq} is singular, since the coefficient of its highest
    derivative vanishes, and
    we must supply the analyticity condition
    \begin{equation}\label{initial_cond}
      \left.u^*\right.'(0)=-\frac{2N}{A_d(N+2)}u^*_0(1+u^*_0),
      \ \ \ u^*_0\equiv u^*(0),
    \end{equation}
    which reduces the possible solutions to a one parameter set.

\begin{figure}[t]
      \centerline{\psfig{file=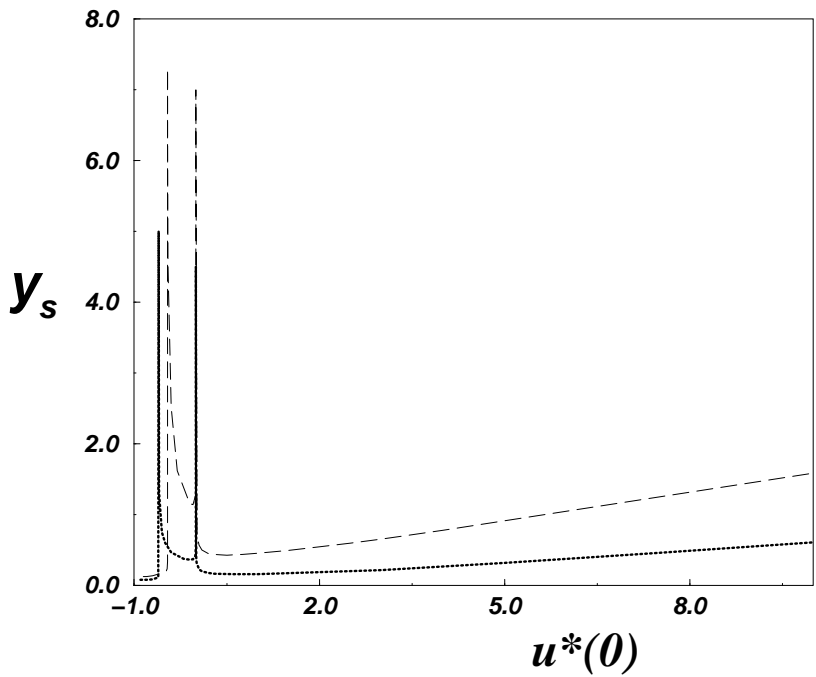,width=\textwidth}}
      \caption{Point $y_s$ where the singularity is encountered as a function
	of the FP initial condition $u^*(0)$, for $N=1$ (dashed line) and
	$N=4$ (solid line), in $d=3$.} 
	\label{fig__sing}
    \end{figure}

    A careful study shows, however, that for any initial value
    \(
      u^*_0\le-1
    \)
    the solution remains negative for any $y$,
    resulting in an unbounded potential.
    We, thus, end up with
    \(
      u^*_0>-1
    \)
    as only possible values.
    Furthermore, both analytical and numerical considerations show that
    solutions of Eq.\
    \ref{FP_eq} end up with a singularity
    \begin{equation}\label{sing}
      u(y)\begin{array}[t]{c}\sim\\\scriptstyle y\nearrow y_s\end{array}
	\frac{A_d}{N(d-2)y_s}\ln(y_s-y)
    \end{equation}
    for all but a finite number of
    \(
      u^*_0
    \).
    In Fig.\ \ref{fig__sing}
    we plot the singular point
    \(
      y_s
    \)
    as a function
    of the initial condition
    \(
      u^*_0
    \)
    for different values of $N$ at $d=3$.
    For
    \(
      d<4
    \)
    only two initial values are allowed, one
    corresponding to the Gaussian Fixed Point (GFP) and another one to
    the Heisenberg Fixed Point (HFP)\footnote{In fact, there is a third one,
    $u^*_0\rightarrow+\infty$.
    It corresponds to the Trivial
    Fixed Point with zero correlation length (infinite mass).}.
    We study each in turn.

    The GFP is 
    \begin{equation}\label{gfp}
      u^*(y)=0.
    \end{equation}
    Its eigenoperators
    are solutions of the linearized version of Eq.\ \ref{FP_eq}
    around $u=0$,
    \begin{equation}\label{eigen_gfp}
      \lambda g=\frac{A_d}{N}[3g'+2yg''+(N-1)g']+(2-d)yg'+2g,
    \end{equation}
    with
    \(
      g(y)
    \)
    the derivative of the scaling operator and
    \(
      \lambda
    \)
    its eigenvalue.
    For the sake of convenience, we define
    \begin{equation}\label{redef}
      z\equiv\frac{(d-2)Ny}{2A_d}
    \end{equation}
    to obtain
    \begin{equation}\label{laguerre_eq}
      zg''(z)+(1+\frac{N}{2}-z)g'(z)+\frac{2-\lambda}{d-2}g(z)=0,
    \end{equation}
    which is the generalized Laguerre equation.
    Polynomial solutions select $\lambda$ to be
    \begin{equation}\label{cr_exp_gfp}
      \lambda=2-n(d-2)
    \end{equation}
    with $n$ a non-negative integer, and, in such a case,
    \begin{equation}\label{laguerre_pol}
      g(y)=L_n^{(N/2)}({\scriptstyle\frac{N(d-2)}{2A_d}}y)
    \end{equation}
    with
    \(
      L_n^{(N/2)}(z)
    \)
    the generalized Laguerre polynomial
    \cite{AS}.
    Critical
    exponents are the Gaussian ones,
    which allow to define massive interacting QFTs in the vicinity of
    this FP.

    The special case of
    \(
      N=1
    \)
    was previously studied in Ref.\ \cite{HH} in terms of the variable
    \(
      x\equiv\sqrt{y}
    \).
    The solutions are the Hermite polynomials of odd degree,
    \(
      H_{2n+1}(\sqrt{z})
    \).
    Indeed \cite{AS},
    \begin{equation}\label{n1_eigen_gfp}
      L_n^{(1/2)}(z)=\frac{(-1)^n}{n!2^{2n+1}\sqrt{z}}H_{2n+1}(\sqrt{z}).
    \end{equation}
    On the other hand, for
    \(
      N\rightarrow\infty
    \)
    the eigenvectors simplify to
    \begin{equation}\label{large_gfp_eigv}
      g(y)=
	  \left(-\frac{(d-2)N}{2A_d}\right)\frac{1}{n!}
	    \left(y-\frac{A_d}{d-2}\right)^n+\ot(1),
    \end{equation}
    a result which will be derived independently in Section
    \ref{SECT__LargeN}.
    A direct diagrammatic computation of the above results
    is left to Appendix \ref{SECT__App_Gauss}.

    The HFP has to be studied numerically.
    From the asymptotic behaviour of Eq.\ \ref{FP_eq}, it is, for
    $y\rightarrow\infty$,
    \begin{eqnarray}\label{hfp_asymp}
      u(y)&\sim&y^{\frac{2}{d-2}}
	\left[B-\frac{2A_d}{d(d-2)}y^{-\frac{d}{d-2}}\right. \\\nonumber
	&&\left.\mbox{}+\frac{2A_d}{(d^2-4)}
	    \left(1-\frac{4}{(d+2)N}\right)y^{-\frac{d+2}{d-2}}\right],
    \end{eqnarray}
    with $B$ an a priori unknown numerical constant.
    Imposing this dependency for large enough values of $y$, together
    with the consistency condition at
    \(
      y=0
    \),
    selects
    the true FP out of the whole bunch
    of non-analytical solutions.

\begin{figure}[t]
      \centerline{\psfig{file=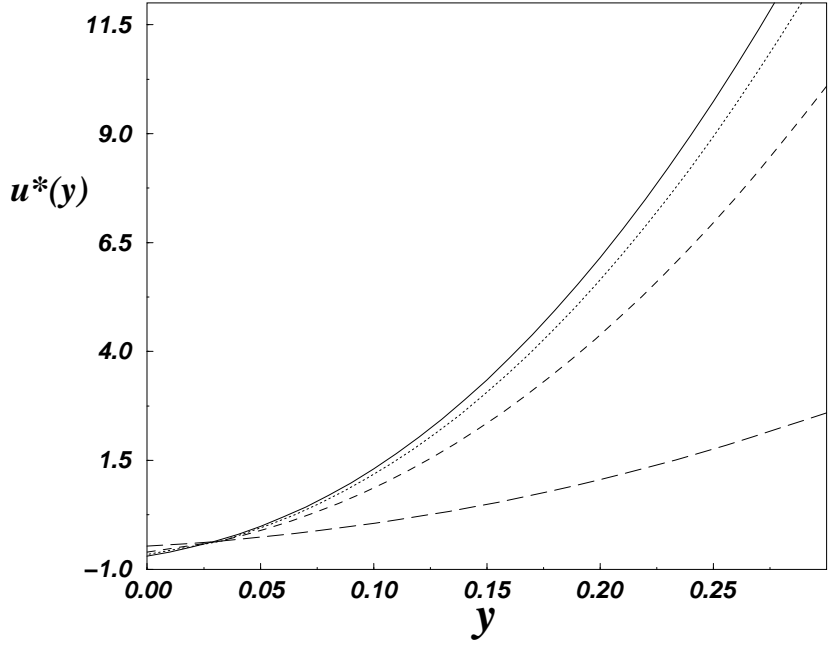,width=\textwidth}}
      \caption{Solutions corresponding to the Heisenberg Fixed Point in
        $d=3$.  The
	long-dashed line corresponds to $N=1$, the dashed one to $N=4$,
	the dotted one to $N=100$ and the solid one
	to $N=\infty$.}\label{fig__FP}
    \end{figure}

    Recall, however,
    that Eq.\ \ref{FP_eq} is singular at the origin.  This makes it
    difficult to numerically integrate the equation from large $y$
    {\em towards} the origin, although it is
    quite easy to integrate it {\em from} the origin.
    We circumvent this difficulty by
    shooting to
    a fitting point \cite{NR}.  That is, we integrate forward to some
    point $y_0$ from a tiny value of
    $y$, satisfying the condition in Eq.\ \ref{initial_cond}.
    Which tiny value to take is immaterial as long as one
    checks it is less than the allowed errors.
    At $y_0$ we compare it with the backward 
    integration from a large value where
    the asymptotic condition is imposed.  In this way, the constants $u^*_0$
    and $B$ from Eqs.\ \ref{initial_cond}, \ref{hfp_asymp} are fixed.
    The precise value of $y_0$, of course, does not matter, but we found it
    efficient to take values close to $\sim 0.05$,
    where the potential is known to
    have a minimum in the exact solution at $N\rightarrow\infty$.
    Plots for some values of
    $N$ at $d=3$ are shown in Fig.~\ref{fig__FP},
    where they are compared with
    the corresponding large $N$ solution.
    
    Linearizing Eq.\ \ref{FP_eq} around
    the HFP, its eigenvectors are found to fulfil
    \begin{eqnarray}\label{eigen_hfp}
      \lambda g&=&\frac{A_d}{N}
	\left[\frac{3g'+2yg''}{1+u^*+2y\left.u^*\right.'}
	  -\frac{(3\left.u^*\right.'+2y\left.u^*\right.'')(g+2yg')
	    }{(1+u^*+2y\left.u^*\right.')^2}\right.  \nonumber\\
	  &&\left.\mbox{}+(N-1)\frac{g'}{1+u^*}
	  -(N-1)\frac{\left.u^*\right.'g}{(1+u^*)^2}
	\right]
        +(2-d)yg'+2g,
    \end{eqnarray}
    which can be solved using similar techniques.
    In general, solutions will grow exponentially,
    \begin{equation}\label{evec_grow}
      g(y)\sim\exp\left(\frac{d^2-4}{2dA_d}NBy^{d/(d-2)}\right),
    \end{equation}
    whereas for special values of
    \(
      \lambda
    \)
    they are much smoother,
    \begin{eqnarray}\label{hfp_evec_asymp}
      \lefteqn{g(y)\sim y^{\frac{2-\lambda}{d-2}}
	\left\{1
	  +\frac{\lambda A_d}{d(d-2)B}
	    \left[1-\frac{2\lambda}{N(d+2)}\right]y^{-\frac{d}{d-2}}\right.}
	\\\nonumber
	  &&\left.\mbox{}-\frac{(\lambda+2)A_d}{(d^2-4)B^2}
	    \left[1-\frac{4(d+2)+2\lambda(d-2)}{N(d+2)^2}\right]
	    y^{-\frac{d+2}{d-2}}
	\right\}.
    \end{eqnarray}
    This asymptotic behaviour, together with an arbitrary
    normalization of eigenvectors, leads to the quantization of $\lambda$.
    The results for $\nu$ and $\omega$ are summarized in Table \ref{tab__exp},
    where $\nu=1/\lambda_0$ and $\omega=-\lambda_1$,
    $\lambda_0$, $\lambda_1$ being the first (the relevant) eigenvalue and
    the second (the first irrelevant) one, respectively.
    \begin{table}[t]
      \centerline{
      \begin{tabular}{|r|l|l||r|l|l||r|l|l|}
        \multicolumn{1}{c}{$N$} & \multicolumn{1}{c}{$\nu$}    &
  			          \multicolumn{1}{c}{$\omega$} &
        \multicolumn{1}{c}{$N$} & \multicolumn{1}{c}{$\nu$}    &
			          \multicolumn{1}{c}{$\omega$} &
        \multicolumn{1}{c}{$N$} & \multicolumn{1}{c}{$\nu$}    &
			          \multicolumn{1}{c}{$\omega$} \\\hline
         0 & 0.6066 & 0.5432 &  7 & 0.9224 & 0.8876 &   50 & 0.9895 & 0.9861 \\
         \hline
         1 & 0.6895 & 0.5952 &  8 & 0.9323 & 0.9028 &   60 & 0.9913 & 0.9884 \\
         \hline
         2 & 0.7678 & 0.6732 &  9 & 0.9401 & 0.9145 &   70 & 0.9925 & 0.9901 \\
         \hline
         3 & 0.8259 & 0.7458 & 10 & 0.9462 & 0.9238 &   80 & 0.9935 & 0.9941 \\
         \hline
         4 & 0.8648 & 0.8007 & 20 & 0.9736 & 0.9639 &   90 & 0.9942 & 0.9924 \\
         \hline
         5 & 0.8910 & 0.8396 & 30 & 0.9825 & 0.9764 &  100 & 0.9948 & 0.9931 \\
         \hline
         6 & 0.9092 & 0.8673 & 40 & 0.9869 & 0.9825 & 1000 & 0.9995 & 0.9994 \\
         \hline
      \end{tabular}}
      \caption{The critical exponents $\nu$ and $\omega$ for different
	values of $N$ and $d=3$.  All digits are significant.  The
	$N\rightarrow\infty$ result is $\nu=\omega=1$.}\label{tab__exp}
    \end{table}
    With nearly the same ease next irrelevant operators can also be studied.
    In Table \ref{tab__irrel} we present our results.
    \begin{table}[t]
      \centerline{\begin{tabular}{|c|r|r||c|r||c|r||r|r|}
        \multicolumn{1}{c}{$N$}
        & \multicolumn{1}{c}{$\lambda_2$} & \multicolumn{1}{c}{$\lambda_3$} &
        \multicolumn{1}{c}{$N$}
	& \multicolumn{1}{c}{$\lambda_2$} &
        \multicolumn{1}{c}{$N$}
	& \multicolumn{1}{c}{$\lambda_2$} &
        \multicolumn{1}{c}{$N$}
	& \multicolumn{1}{c}{$\lambda_2$} \\\hline
	\multicolumn{1}{|c|}{1} & -2.8384
	& -5.1842 & 5 & -2.8873 & 10 & -2.9426 &  60 & -2.9917 \\\hline
	\multicolumn{1}{|c|}{2} & -2.8348
	& -5.1100 & 6 & -2.9036 & 20 & -2.9731 &  70 & -2.9928 \\\hline
	\multicolumn{1}{|c|}{3} & -2.8482
	& -5.0582 & 7 & -2.9167 & 30 & -2.9827 &  80 & -2.9938 \\\hline
	\multicolumn{1}{|c|}{4} & -2.8680
	& -5.0286 & 8 & -2.9273 & 40 & -2.9873 &  90 & -2.9945 \\\hline
	\multicolumn{3}{c|}{}
	          & 9 & -2.9357 & 50 & -2.9899 & 100 & -2.9951 \\\cline{4-9}
      \end{tabular}}
      \caption{The third and fourth eigenvalues, $\lambda_2$, $\lambda_3$,
        for different $N$s at $d=3$.
        All digits are significant.  The $N\rightarrow
        \infty$ result is $\lambda_2=-3$, $\lambda_3=-5$.}\label{tab__irrel}
    \end{table}

    The $N=1$ case was first computed in Ref.\ \cite{HH}, with results
    $\nu=0.687(1)$ and $\omega=0.595(1)$, and afterwards in Ref.\
    \cite{tim:spur}.  Our numbers are compatible with the
    former and coincide with the latter.

\begin{figure}[t]
      \centerline{\psfig{file=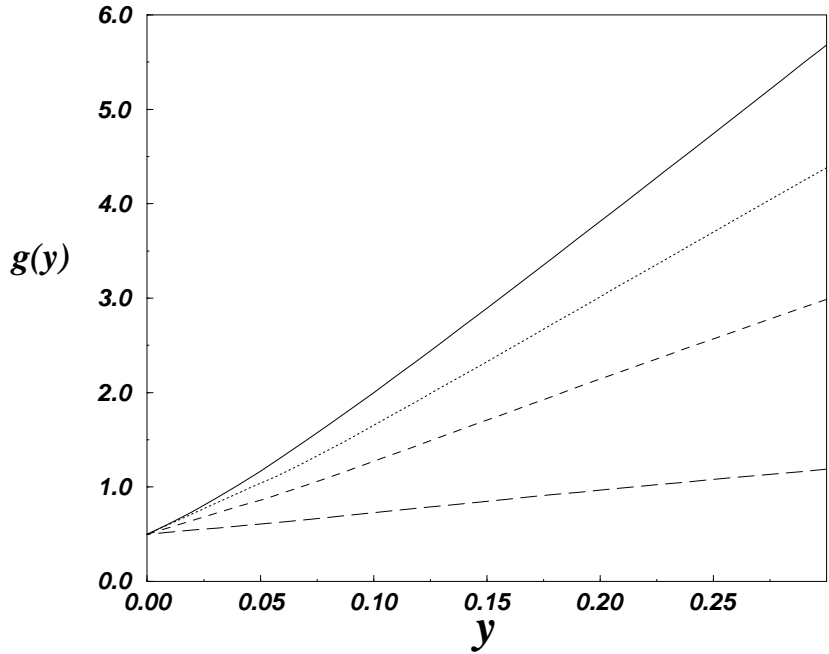,width=\textwidth}}
      \caption{The first eigenoperator, normalized to $g(0)=1/2$, for
	$N=1$ (long-dashed line), $N=4$ (dashed line), $N=100$ (dotted
	line), $N=\infty$ (solid line), at $d=3$.}\label{fig__eigen}
    \end{figure}

    Eigenoperators are obtained
    as an additional bonus.  As a
    matter of example, we plot in Fig.\ \ref{fig__eigen}
    the first
    eigenvector, normalized to
    \(
      g(0)=1/2
    \),
    corresponding
    to different values of $N$ at $d=3$.

\begin{figure}[t]
      \centerline{\psfig{file=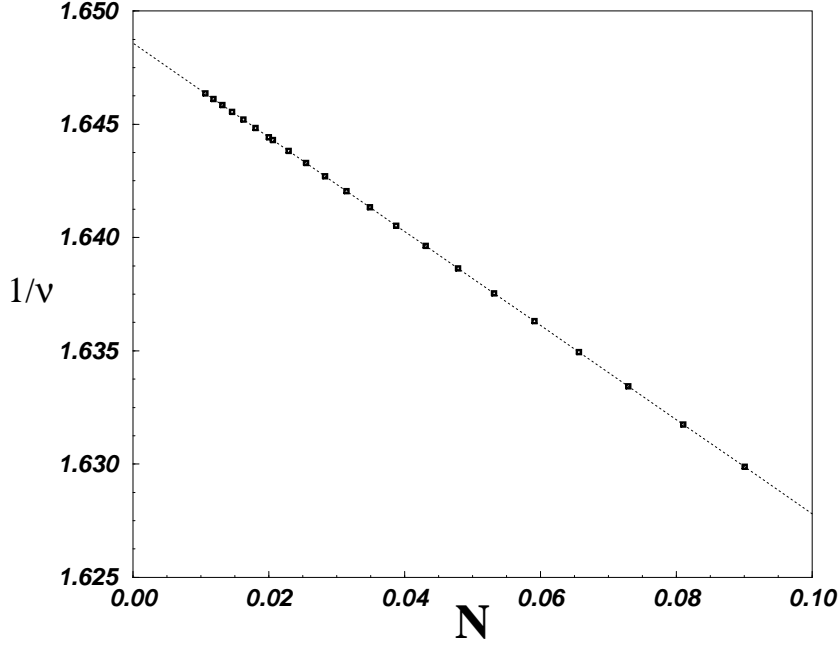,width=\textwidth}}
      \caption{The first eigenvalue for small values of $N$ at $d=3$.
      The dots are
      the calculated numbers, the straight line is fitted to extrapolate
      to $N=0$.}\label{fig__N0}
    \end{figure}

    The
    \(
      N\rightarrow0
    \)
    limit deserves a special remark,
    as the RG equation \ref{proj_eq} is singular at
    \(
      N=0
    \).
    Nevertheless,
    one may compute its eigenvalues in an indirect way.
    Taking
    values of $N$
    between
    \(
      0.1
    \)
    and
    \(
      0.01
    \),
    one finds that they perfectly fit a straight line,
    with no
    sign at all of subleading corrections.
    Critical exponents reported in Table \ref{tab__exp}
    are just their extrapolation to $N=0$.
    In Fig.\ \ref{fig__N0} we plot the first eigenvalue for
    \(
      0.01<N<0.1
    \)
    together with its linear extrapolation.

    This apparent singularity
    at $N=0$ is, however, due to the rescaling in Eq.\ \ref{norm_N}.
    Had we defined the partition function without explicit factors of $N$,
    the $A_d/N$ coefficient in RG Eq.\ \ref{proj_eq} would be
    simply $A_d$, allowing
    a direct analysis of the $N\le0$ values.
    We just decided to stick to our
    normalization.
  \section{Comparison with other equations}\label{SECT__Comparison}
    In this section we comment on other ERG equations
    and compare with known results.

    Apart from the Wegner-Houghton equation studied above, one of the oldest
    ERG equations appeared in Ref.\ \cite{WILS}.  It was further
    modified by Polchinski \cite{Polchinski} to put it in the form we now use.
    The regulator is chosen in such a way that every propagator is multiplied
    by a function
    \(
      K(p^2)
    \)
    which is analytic everywhere in the finite complex plane, vanishes faster
    than any polynomial for large real values of its argument and it is
    standardly normalized to
    \(
      K(0)=1
    \)
    \cite{Ball}.
    The evolution of the action takes the form
    \begin{eqnarray}\label{Pol_eq}
      \dot S&=&\frac{\delta S}{\delta\phi}\cdot K'\cdot
            \frac{\delta S}{\delta\phi}
          -\frac{1}{N}\,\mbox{tr}\left(K'\cdot
              \frac{\delta^2S}{\delta\phi\delta\phi}\cdot\right)
          -2\phi\cdot p^2K^{-1}K'\frac{\delta S}{\delta\phi}\nonumber\\
        &&\mbox{}+dS+\frac{2-d-\eta(t)}{2}\,
          \phi\cdot\frac{\delta S}{\delta\phi}
          -\phi\cdot\left(p\cdot\frac{\partial'}{\partial p}\right)\frac{
              \delta S}{\delta\phi}+\mbox{const},
    \end{eqnarray}
    where a dot means both summation over discrete variables and integration
    over continuous ones.
    As the non-linearities are limited to quadratic terms,
    the projected equation presents a nice form quite amenable for numerical
    analysis,
    \begin{equation}\label{Pol_proj}
      \dot u=\frac{2y}{N}u''+\left(1+\frac{2}{N}+(2-d)y-2yu\right)u'+(2-u)u,
    \end{equation}
    where a convenient rescaling of the potential and the field variable $y$
    has been made.
    It has been previously studied for $N=1$, using the
    variable $x\equiv\sqrt{2y}$, in Ref.\ \cite{BHLM}.

    The GFP is also $u=0$, with its eigenoperators satisfying
    Eq.\ \ref{laguerre_eq} after the rescaling $z\equiv\frac{N}{2}(d-2)y$.
    The solution $u=2$ corresponds to the Trivial Fixed Point,\footnote{To
    substantiate it, consider a potential consisting only of a mass term,
    $u=\frac{2re^{2t}}{1+re^{2t}}$.  Note that at first order in $r$
    (or, alternatively, for $t\rightarrow-\infty$) it coincides with the
    relevant mass operator from the GFP, $u=0$, whereas
    $\frac{2re^{2t}}{1+re^{2t}}\rightarrow2$ for $t\rightarrow\infty$.
    Furthermore, it is easily proved that all linear
    deviations from $u=2$ are irrelevant.}
    also known as High Temperature Fixed Point.

    The analysis of the HFP is analogous to the Wegner-Houghton one.
    There are singular solutions which behave
    \begin{equation}\label{Pol_sing}
      u\sim\frac{2}{N}\frac{1}{y_s-y},
    \end{equation}
    whereas the true FP must fulfil the conditions
    \begin{eqnarray}\label{Pol_cond}
      \left.u^*\right.'(0)&=&\frac{N}{N+2}u^*(0)(u^*(0)-2),\nonumber\\
      u^*(y)&\begin{array}[t]{c}\sim\\
              \scriptstyle y\rightarrow\infty
	    \end{array}&
	2-By^{-\frac{2}{2+d}}
    \end{eqnarray}
    with $B$ a non-vanishing constant.  Again we can shoot to a fitting point
    to obtain the FP. 
    Eigenvalues $\lambda$
    are obtained from the linearized form of Eq.\ \ref{Pol_proj}
    imposing the eigenvectors to be
    \(
      g\sim y^{-(\lambda+2)/(d+2)}
    \)
    for
    \(
      y\rightarrow\infty
    \).
    Some of them are shown in Tables \ref{tab__nu}, \ref{tab__omega}.
    We leave to Appendix \ref{SECT__App_LargeN} the study of its large $N$
    limit.

    Another quite used equation is based on the evolution of an
    infrared-regulated
    one-particle-irreducible effective action \cite{tim:ijmp},
    \begin{equation}\label{1PI}
      \frac{\partial}{\partial\Lambda}\Gamma[\varphi]=-\frac{1}{2}\mbox{tr}
        \left[
          K^{-1}\frac{\partial K}{\partial\Lambda}\cdot
          \left(
            1+K\cdot\frac{\delta^2\Gamma}{\delta\varphi\delta\varphi}
	  \right)^{-1}
	\right],
    \end{equation}
    where $K$ is some regulating function and
    \(
      \Gamma[\varphi]
    \)
    is the effective action minus the kinetic term
    \(
      \frac{1}{2}\varphi\cdot K^{-1}\cdot\varphi
    \).
    It presents some
    attractive features: in the limit where the IR cutoff vanishes (which
    corresponds to the continuum limit) it is directly observable, as it is
    the equation of state of the system.  Moreover, it presents a natural
    expansion which can be directly mapped to the usual loop expansion of
    perturbation theory \cite{Wet}.
    Various regulators have been used in connection
    with this equation.
    If a sharp momentum cutoff is chosen, its LPA coincides with
    that of Wegner-Houghton's.  On the other hand, it has been extensively
    worked out using
    the special polynomial cutoff introduced
    in Ref.\ \cite{tim:ON}, which has the advantage of preserving
    some symmetries of the exact equation through the derivative
    expansion.  This equation is much harder to numerically analyze
    than the above two ones.
    We do not discuss it further, but only
    quote the results (Tables \ref{tab__nu}, \ref{tab__omega})
    and refer the reader to the original literature
    \cite{tim:ON,tim:Dub}.

    To assess the systematic errors involved in such calculations 
    would require to compute corrections to the LPA. Instead, we compare
    our results with some derived from other techniques.

    There are accurate determinations for $\nu$ from a variety of methods
    such as $\varepsilon$ expansion \cite{EPS}, Monte Carlo 
    Renormalization Group \cite{GUP,Crit_mcrg}, Monte Carlo simulations
    \cite{MAD,Crit_MC} and Strong-Coupling series \cite{Crit_SC}. 
    Unfortunately, estimates for $\omega$ are not so precise.
    A comparison of our results from Wegner-Houghton and Polchinski equations
    with those from the IR-cutoff effective action and other existing
    estimates are shown in Tables \ref{tab__nu}, \ref{tab__omega}.
    \begin{table}[t]
      \centerline{\begin{tabular}{|c||l|l|l|l|}
        \multicolumn{1}{c}{$N$}           &
        \multicolumn{1}{c}{(1)}           &
        \multicolumn{1}{c}{(2)}           &
        \multicolumn{1}{c}{(3)}           &
        \multicolumn{1}{c}{(4)}
         \\\hline
        0 & 0.6066 & \multicolumn{2}{c|}{} & 0.5880(15)  \cite{EPS}
         \\\hline
        1 & 0.6895 & 0.6496 & 0.6604 \cite{tim:ON}  & 0.624(1)(2) \cite{GUP}
         \\\hline
	2 & 0.7678 & 0.7082 & 0.73   \cite{tim:Dub} & 0.6721(13)  \cite{MAD}
	 \\\hline
	3 & 0.8259 & 0.7611 & 0.78   \cite{tim:Dub} & 0.7128(14)  \cite{MAD}
	 \\\hline
	4 & 0.8648 & 0.8043 & 0.824  \cite{tim:Dub} & 0.7525(15)  \cite{MAD}
	 \\\hline
      \end{tabular}}
      \caption{Critical exponent $\nu$, at $d=3$,
        for different values of
        $N$, using different RG equations (all digits are significant):\
        (1) Wegner-Houghton, (2) Polchinski, (3) effective action;
        together with accurate determinations (4).}
      \label{tab__nu}
    \end{table}
    \begin{table}[t]
      \centerline{\begin{tabular}{|c||l|l|l|l|}
        \multicolumn{1}{c}{$N$}           &
        \multicolumn{1}{c}{(1)}           &
        \multicolumn{1}{c}{(2)}           &
        \multicolumn{1}{c}{(3)}           &
        \multicolumn{1}{c}{(4)}
         \\\hline
        0 & 0.5432 & \multicolumn{2}{c|}{} & 0.80(4)   \cite{EPS}
         \\\hline
        1 & 0.5952 & 0.6557 & 0.6285 \cite{tim:ON}  & 0.85(5)   \cite{GUP}
         \\\hline
	2 & 0.6732 & 0.6712 & 0.66   \cite{tim:Dub} & 0.780(25) \cite{EPS}
	 \\\hline
	3 & 0.7458 & 0.6998 & 0.71   \cite{tim:Dub} & 0.800(25) \cite{EPS}
	 \\\hline
	4 & 0.8007 & 0.7338 & 0.75   \cite{tim:Dub} & \multicolumn{1}{c|}{?}
	 \\\hline
      \end{tabular}}
      \caption{Critical exponent $\omega$, at $d=3$,
        for different values of
        $N$, using different RG equations (all digits are significant):\
        (1) Wegner-Houghton, (2) Polchinski, (3) effective action;
        together with existing estimates (4).}
      \label{tab__omega}
    \end{table}

    Critical indices from Wegner-Houghton equation are $10-15 \%$ off
    for $N<10$, except at $N=0$, which is quite accurate ($\sim 3 \%$). In 
    the interval $10<N<100$, they are
    off by a few per cent, and for $N>100$, the
    errors are less than one per mile. Other ERG perform slightly
    better, specially Polchinski's equation in which critical indices
    are a few per cent off at most. This should not come as a surprise,
    since in spite of its conceptual clarity, the sharp momentum
    cut-off leads to bad locality properties 
    of the renormalized actions, being then
    very much sensitive to truncations. In order for the projected ERG
    equation to deliver more precise outcomes, one should find
    RG blockings with the property that renormalized actions are as much
    local as possible,
    as it is achieved in the lattice regularization
    with blockings in real space \cite{WIL_BELL}.
 
  \section{$N\rightarrow\infty$}\label{SECT__LargeN}
    Wegner and Houghton \cite{WH}
    already studied the limit $N\rightarrow\infty$ of their equation,
    encountering both the GFP and the HFP.  They also
    briefly mentioned the possibility
    of further FPs.  Our aim
    is to discuss in certain detail those not previously studied and
    present a general framework to
    deal with this kind of large $N$ equations.
    An application of these methods for the Polchinski equation is given
    in Appendix \ref{SECT__App_LargeN}.

    The projected RG equation becomes first order in the
    $N\rightarrow\infty$ limit,
    \be\label{LargeN_eq}
      \dot u=A_d\frac{u'}{1+u}+(2-d)yu'+2u,
    \ee
    which is easily solved for $2<d<4$
    once we invert $u$ and consider $y(u,t)$, 
    \be\label{lin_LargeN_eq}
      \dot y=-2uy'-(2-d)y-\frac{A_d}{1+u}.
    \ee
    The prime now stands for derivative with respect to $u$.  Its
    general solution is
    \be\label{Gen_sol}
      y(u,t)=e^{(d-2)t}h(e^{-2t}u)+f(u),
    \ee
    with an arbitrary function $h(z)$ and
    \be\label{LargeN_Gen_sol}
    f(u)=A_d\left(\frac{1}{d-2}+u\int_0^1dz\frac{z^{3-d}}{1+uz^2}\right),
    \ee
    which is analytic for $u>-1$.  Although derived differently,
    this is the form of the ERG equation used in Refs.\ \cite{MA,david}
    for $d=3$.

    Returning to Eq.\ \ref{LargeN_eq},
    FPs verify
    \be\label{LargeN_eq_rear}
      u'=\frac{2u}{(d-2)y-\frac{A_d}{1+u}},
    \ee
    which can be singular when the denominator of the rhs vanishes. But,
    if this is the case, then analyticity of the FP would imply $u=0$, and,
    therefore, these would-be singular points reduce to a unique one,
    \be
    {\tilde y}\equiv \frac{A_d}{d-2}, \ \ \  u(\tilde y)=0.
    \ee
    Furthermore, we can Taylor expand the FP around $\tilde y$,
    \be\label{LargeN_Taylor}
    u(y)=a(y-\tilde y)+\sum_{k=2}^{\infty}a_k(y-\tilde y)^k,
    \ee
    an expansion that will be used throughout.
    
    Eigenoperators are
    \be\label{LargeN_eigen}
      g_{\lambda}(y)=\exp \int_0^ydz \,t_{\lambda}(z), 
    \ee
     with
    \be\label{LargeN_sing}
      t_{\lambda}(y)=\left[\frac{A_du'}{(1+u)^2}+\lambda-2\right]
                     \left[\frac{A_d}{1+u}-(d-2)y\right]^{-1}. 
    \ee
    Eq.\ \ref{LargeN_Taylor} fixes the only singularity of $t_{\lambda}(y)$ 
    to be a pole at $\tilde y$, with residue 
    \be\label{LargeN_res}
    \frac{1}{1+a\tilde y}\left(\frac{\lambda-2}{2-d}-a\tilde y\right).
    \ee
    For eigenoperators to be single-valued, the residue must be a nonnegative
    integer $n$, which fixes
    \be\label{LargeN_value} 
    \lambda_n=2+(2-d)[a\tilde y +n(1+a\tilde y)].
    \ee
    Note how critical properties of the system are completely determined
    by the behaviour of the FP around this special point $\tilde y$.

    After those general considerations let us study the different FPs in turn.
    The easiest one is the GFP, which corresponds to
    \be\label{LargeN_gaus}
      u(y)=0 \Rightarrow a=0 \Rightarrow \lambda_n=2-n(d-2)
    \ee
    and, from Eq.\ \ref{LargeN_eigen},
    \be
    g(y)=(y-\tilde y)^n,
    \ee
    a result previously obtained in Eq.\ \ref{large_gfp_eigv}.

    The HFP corresponds to $h(z)=0$ in Eq.\ \ref{Gen_sol},
    \be\label{LargeN_HFP}
      y(u)=f(u)=
        A_d\left(\frac{1}{d-2}+u\int_0^1dz\frac{z^{3-d}}{1+uz^2}\right).
    \ee
    In this case, $a=\frac{4-d}{A_d}$.  Hence,
    \be\label{LargeN_Hei}
      \lambda_n=d-2-2n,
    \ee
    which are the critical exponents of the spherical model. 
    Eigenvectors
    are obtained plugging Eq.\ \ref{LargeN_HFP} in Eq.\ \ref{LargeN_eigen}.
    Note that in the limit $d\rightarrow4$, the HFP merges with the GFP,
    as can be seen both from Eqs.\ \ref{LargeN_HFP} and \ref{LargeN_Hei}.

    However, contrary to the finite $N$ case, there are more FP.
    They correspond to
    $ e^{(d-2)t}h(e^{-2t}u)$ becoming t-independent,
    \be\label{LargeN_FPBMB}
       y(u)-\tilde y=C u^{\frac{d-2}{2}}+A_du\int^1_0dz
       \frac{z^{3-d}}{1+uz^2},
    \ee 
    with $C$ a, so far, arbitrary constant. The behaviour around $\tilde y$,
    \be
    u(y)=\left( \frac{y-\tilde y}{C} \right)^{\frac{2}{d-2}}+
    \ot\left((y-\tilde y)^{\frac{4}{d-2}}\right),
    \ee
    defines for $2/(d-2)=n$, $ n=2,3,\ldots$
    a line of FP labelled by the parameter $C$.
    Since $a=0$,
    critical indices coincide with those of the GFP, for all allowed values
    of $C$.

    Before getting into a detailed analysis, let us state that,
    for any allowed $C$,
    these FPs are reached by Eq.\ \ref{Gen_sol}, at $t\rightarrow
    \infty$, from an initial potential
    \be\label{potential}
      V(y)=\frac{C}{2(n+1)}\left(\frac{y-nA_d/2}{C}\right)^{n+1},
    \ee
    which shows that those FPs encode the large distance properties of local
    actions.
    Note that
    $C=+\infty$
    corresponds to the GFP. 
    
    To proceed further, we should distinguish between $n$ odd and
    $n$ even. 
    The FP in the $n$ odd case is
    \be
      y=Cu^{\frac{1}{n}}+A_d\left(\frac{n}{2}+u
      \int_0^1dt\frac{t^{1-2/n}}{1+ut^2}\right).
    \ee
    The admissible values of $C$ are $0<C\le\infty$.  Although the HFP
    corresponds to $C=0$, it does not belong to this line of FPs.  This
    is not so surprising in view of Eq.\ \ref{potential}:\ critical 
    properties are not continuous in $C$ at $C=0$.
    Note, however, that these singularities have nothing to do with
    RG transformations:\ $C$ labels different initial conditions for
    the RG transformations, but once one $C$ is chosen, it is not changed
    along the flow.

    The FP in the $n$ even case comes from matching the two branches
    of $y(u)$ at $\tilde y$, 
    \be\label{LargeN_BMB}
      y=\left\{
      \begin{array}{lc} 
      -Cu^{\frac{1}{n}}+f(u) & 0 \le y \le \tilde y\\
      +Cu^{\frac{1}{n}}+f(u) & y \ge \tilde y
      \end{array} 
      \right.
    \ee
    Now the line of FPs extend from $C=+\infty$ up to the end point
    \be\label{c_admis}
      C=\frac{A_d \pi}{2 \sin\frac{\pi}{n}},
    \ee
    which appears as the existence condition of the FP at $y=0$.
 
    The case $C=\frac{A_d \pi}{2\sin\frac{\pi}{n}}$ deserves
    further considerations because
    the 
    FP is not analytical at $y=0$.  It behaves
    like $u(y)=\frac{nA_d}{2(n+1)}\frac{1}{y}+{\cal O}(1)$, and
    its eigenvectors like,
    \be
      g_k(y)\sim y^{-\frac{(n+1)(n+1-k)}{n(n-1)}}.
    \ee
    The case $n=2$ ($d=3$) were previously studied
    in Ref.\ \cite{david} and these features were identified as the RG
    structures which make the existence of the BMB phenomenon \cite{BMB}
    possible. That is, the possibility that at
    $C=\frac{A_3 \pi}{2\sin\frac{\pi}{2}}$, the FP acts as an ultraviolet
    stable one, allowing to define a new continuum limit other
    than the Gaussian and Heisenberg FPs.  In fact one that
    has mass generation and spontaneous symmetry breaking of scale 
    invariance accompanied with the appearance of a dilaton. What we
    now see, is that all RG properties that make this phenomenon to
    appear at $d=3$, are present for all the rest of $n$ even values,
    hence a similar scenario is likely to take place.
    To elucidate these issues would require a further
    analysis of the effective
    potential at the vicinity of this peculiar FP.
    This is, however, beyond the scope of this paper.

    The preceding considerations have shown the existence of a line
    of FPs starting at the GFP and including it. This scenario is
    only feasible provided that the GFP contains marginal non-redundant
    operators. This restricts the dimensions of space (space-time) to be
    \be
     d=2+\frac{2}{n},\ \ \ n=2,3,\ldots
    \ee
    which gives a nice physical interpretation of the dimensions
    previously derived from analyticity considerations. Thus, in the
    large $N$ limit, the marginal operator becomes completely
    marginal and a line of inequivalent FP appears, though they
    have the critical exponents of the GFP. This might be a general
    feature of this limit.

    Before closing this section we would like to remark that, although
    our analysis
    of Eq.\ \ref{lin_LargeN_eq} is restricted 
    to dimensions $2<d<4$,
    a similar study in other dimensions is possible along the same
    lines.  For instance, in
    $d=4$, Eq.\ \ref{Gen_sol} is substituted by
    \be
    y(u,t)=e^{2t}h(e^{-2t}u)+\frac{u}{8\pi^2}t+\frac{1}{16\pi^2}-
    \frac{u}{8\pi^2}\int_0^1dz\,\frac{z}{1+uz^2},
    \ee
    which
    does not have any FP other than the GFP, $u=0$. 
    On the contrary, at $d=2$,
    \be
    y(u,t)=h(e^{-2t}u)-\frac{t}{2\pi}+\frac{1}{4\pi}\ln(1+u),
    \ee
    where a FP appears for
    $h(z)=-\frac{1}{4\pi}\ln(z)$.
    A thorough study for all $d$ would take us too far afield.

  \section{Conclusions}\label{SECT__Conclusions}

    This paper presented a detailed study of Wegner-Houghton equation.
    We have shown that in spite of its intractable appearance, the use of
    the LPA makes it amenable for extracting plenty of valuable 
    information.

    The derivation of the equation has been worked out, with special care
    about the terms that contribute to the blocking.
    The projection has been derived, and explained in which sense it is
    exact in the $N\rightarrow\infty$ limit.

    What we feel remarkable about the ERG methods within the LPA
    is, succinctly;  (i) The qualitative features of the RG at $2<d\le 4$
    are the same as in the
    expected full computation;
    (ii) Eigenvalues and eigenvectors are obtained with ease, both
    relevant and irrelevant ones;  (iii) Numerical results are, at
    this order in the derivative expansion, not as accurate
    as other methods, at least for low enough $N$, but those are
    computationally much harder;  (iv) The $N\rightarrow0$ limit is
    reliable within this framework.

    The LPA of Polchinski equation has been also
    studied, and found that the accuracy is improved.  This
    suggests that a calculation of next order in 
    the momentum expansion may deliver
    quite rigorous results.  Nevertheless, ambiguities appear in connection
    with the determination of the non-vanishing anomalous dimension, which
    are not present at lowest order \cite{BHLM}.  A further study of
    these issues, hence, seems justified \cite{mine}.

    Section \ref{SECT__LargeN} has been
    devoted to the $N\rightarrow\infty$ limit.
    Some existing results, scattered throughout
    the literature, have been reviewed within a general framework, which
    has also been fruitful to study new FPs at critical
    dimensions $d=2+2/n$, as well as
    to deal with other RG equations.  As an example, we have worked out
    the large $N$ limit of Polchinski equation.

    The scope of the paper did not permit
    to consider other subjects of interest.  Among those,
    the study of the projected RG equation \ref{proj_eq}, outside the
    linear approximation, useful to elucidate the topology of the RG
    \cite{HH,SACL};
    the study of the effective potential of the new FPs at 
    $N \rightarrow \infty$, with special emphasis on the singularities
    that we associate to the BMB phenomenon; and a more systematic
    study of Polchinski equation.  We hope that those issues will be
    tackled in a near future.

    Before closing, we would like to come back to our original
    motivation on first-order phase transitions, singling out
    some of our results which have implications for them.
    We have explicitly shown
    in Section \ref{SECT__Projection} that in the large $N$ limit,
    the LPA is exact (see Eq.\ \ref{exact}).  This equation, in the form of
    Eq.\ \ref{Gen_sol}, was shown in Ref.\ \cite{HH} to lead to 
    singularities of the RG whenever two competing minima exist,
    which is the case when first-order phase transitions
    take place.
    Those exact results point that
    the RG trajectories are singular and multi-valued, which is at odds with
    Fundamental Theorems One and Two of Refs.\ \cite{SOK}.
    Nonetheless, one should keep in mind
    that those theorems do not rigorously apply for unbounded spins.
    On the other hand, the derivation of Wegner-Houghton equation
    of Section \ref{SECT__Deriv}
    assume positivity of the eigenvalues of $S''_{q,-q}$,
    at the step of Eq.\ \ref{integ_gaus},
    which, as suggested in Ref.\
    \cite{HH}, may not be the case if metastable states
    do exist.
    A further study on all those subjects is beyond the
    scope of this paper, but we think that they pose interesting
    questions on the conventional wisdom of the RG.
  \section*{Acknowledgments}
    It is a pleasure to acknowledge interest and discussions with
    D. Espriu, J.I. Latorre, S. Leseduarte, T.R. Morris and F. Sala.
    The contribution of J. Herrero and A. Pineda in our discussions is
    also acknowledged.
    J. C. acknowledges H. Osborn and DAMTP, where this work was
    initiated, for kind hospitality.
    A. T. has benefited from a fellowship from
    Generalitat de Catalunya.  This work
    has been supported by grants AEN95-0590 (CICYT) and GRQ93-1047 (CIRIT).

    \bigskip
    \bigskip

    After the completion of this paper, we became aware of the work of
    Ref.\ \cite{JAP}, where a computation of the exponent $\nu$, for
    different values of $N$, is performed, using the LPA with a further
    truncation in the number of fields.  Although this
    additional expansion
    is known to be problematic
    \cite{tim:spur}, their results are in good agreement with ours.

    A recent reference about the large~$N$ limit and the BMB phenomenon
    is~\cite{Moshe}.  We thank M. Moshe for pointing it to us.
  \newpage
  \appendix
  \section{The Gaussian RG Equation}\label{SECT__App_Gauss}

    In this section we work out the GFP from a direct computation
    of the partition function, reobtaining
    in quite a different way expressions already quoted in Section
    \ref{SECT__Results}.  We  use diagrammatic techniques very
    similar to the ones in deriving the approximate recursion
    formula \cite{WILS}.
    We explicitly
    treat the $N=1$ case, but the $N$ arbitrary situation is completely
    analogous.

    We consider the splitting in Eq.\ \ref{split},
    \be\label{app_split}
    \phi_q=\phi^{(0)}_q+\phi^{(1)}_q,
    \ee
    but now $\phi^{(1)}$ accounts for modes with $\frac{1}{2}<q\le1$.
    Then, the dilatation in Eq.\ \ref{dilatation} is
    \be
    k\equiv 2q,\ \ \ \phi_k=2^{-\frac{d+2}{2}}\phi^{(0)}_{q}.
    \ee
    We consider our action as the GFP and the subspace spanned
    by (ultra-)local operators,
    \be\label{app_action}
    S[\phi]={\cal I}+\frac{1}{2}\int_{q}\phi_{q}q^2\phi_{-q}+
      \sum_{l=2,4,\ldots} g^{(l)} \int d^dx \phi^{2l}(x),
    \ee	
    where we explicitly include the identity.

\begin{figure}[t]
      \centerline{\psfig{file=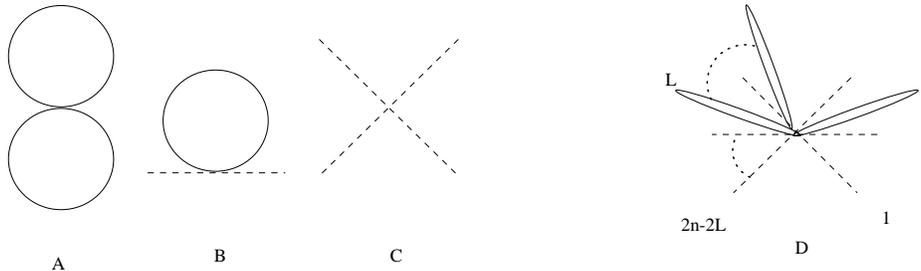,width=\textwidth}}
      \caption{Diagrams to be considered, where solid lines mean
      integrated modes and dashed lines are not.
       A, B, C, are the contributions
      of the quartic vertex to the identity, the mass operator and the
      quartic coupling itself. D is a generic diagram with $L$ loops.
      }\label{fig__apen}
    \end{figure}

    The linear approximation amounts to consider $g^{(l)}$ couplings
    to be infinitesimal. In such a case we can compute perturbatively the
    flow.  In Fig.\ \ref{fig__apen}, the three diagrams (A,B,C) 
    generated from a quartic coupling are shown.
    A generic diagram of a $2n$ operator
    with $L$ loops, like the one in Fig.\ \ref{fig__apen} D,
    gives a contribution to the $2n-2L$ coupling 
    \be\label{app_result}
      \frac{(2n)!c^L}{(2n-2L)!L!2^L}2^{2n-2L+d(1-n+L)},
    \ee
    where $c=\int_{\frac{1}{2}<|q|\le1}\frac{1}{q^2}$.
    The linear RG matrix is upper triangular, with non-vanishing
    matrix elements,
    \be\label{app_rgmat}
      T_{l,k}=\frac{(2k)!c^{k-l}}{(2l)!(k-l)!}2^{3l-k+d(1-l)}.
    \ee
    The main result is that the eigenoperators of the linearized RG 
    matrix are
    \be\label{app_eigen}
      O_{n}(x)=\sum_{l=0}^{n}g^{(l)}_n\phi^{2l}(x)=
      \sum_{l=0}^{n} \frac{(2n)!}{(2l)!(n-l)!}
        \left(\frac{c}{2(2^{2-d}-1)}\right)^{n-l}\phi^{2l}(x),
    \ee
    with eigenvalues  $\Lambda_n\equiv2^{\lambda_n}=2^{(1-n)d+2n}$.
    The proof is by direct substitution,
    \bea
      \sum_{k} T_{l,k}g^{(k)}_n&=&2^{3l-n+d(1-l)}c^{(n-l)}\frac{(2n)!}{(2l)!}
      \sum_{k=l}^{n}\frac{(2^{2-d}-1)^{k-n}}{(n-k)!(k-l)!}
        \nonumber\\
      &=&2^{2n+d(1-n)}\frac{(2n)!}{(2l)!(n-l)!}
        \left(\frac{c}{2(2^{2-d}-1)}\right)^{n-l}
      =\Lambda_n g^{(l)}_n,
    \eea
    where the relation 
    \be
      \sum_{k=l}^{n} \frac{(n-l)!}{(k-l)!(n-k)!}a^k= \sum^{n-l}_{k=0} 
      {n-l \choose k} a^{l+k}={(1+a)}^{n-l}a^l
    \ee
    has been used.
    Eq.\ \ref{app_eigen} may be now
    rewritten as
    \be\label{app_herm}
      O_n(x)=\left(\frac{A_d}{2(d-2)}\right)^n H_{2n}\left({\scriptstyle\sqrt{
	\frac{d-2}{2A_d}}}\,\phi(x)\right),
    \ee
    since $c={\displaystyle \int_{\scriptscriptstyle \frac{1}{2}<|q|\le 1}}
    \frac{1}{q^2}=\frac{A_d}{d-2}(1-2^{(2-d)})$.
    $H_{2n}$ are the Hermite polynomials, which, after using the relation
    \cite{AS}
    \be
      H_{2n}'(x)=4nH_{2n-1}(x),
    \ee
    reduce to the result already quoted in Eq.\ \ref{n1_eigen_gfp}.

  \section{$N\rightarrow\infty$ with Polchinski equation}
    \label{SECT__App_LargeN}

    In this appendix, we sketch the results for the large $N$ limit of
    Polchinski equation.  Our aim is twofold:\ on one hand to show that our
    techniques introduced
    for Wegner-Houghton equation are quite general
    and may be easily applied to other
    RG equations as well; also to prove that the annoying restriction
    $u>-1$ of the former equation is not an essential feature.  This
    last fact would make interesting to repeat the
    large $N$ analysis of Ref.\ \cite{HAS}
    for Polchinski equation.

    At $N\rightarrow\infty$, Eq.\ \ref{Pol_proj} becomes
    \be
      \dot u=[1-(d-2)y-2yu]u'+(2-u)u.
    \ee
    Its general solution is
    \be
      y(u,t)=\frac{e^{(d-2)t}}{(2-u)^2}\,h\left(\frac{ue^{-2t}}{2-u}\right)
        +f(u),
    \ee
    with $h(z)$ an arbitrary function and
    \be
      f(u)=\frac{2}{d-2}\,(2-u)^{-1}+\frac{d}{d-2}\,u(2-u)^{-\frac{d+2}{2}}
        \int_0^1dz\left(\frac{2-uz}{z}\right)^{\frac{d-2}{2}}.
    \ee
    Note that $f(u)$ is analytic
    for\footnote{Recall
      that $u=2$ corresponds here to the Trivial Fixed Point.}
    $u<2$, unlike the
    Wegner-Houghton case, in which analyticity has a lower bound at $u=-1$.

    For $\dot u=0$ the point $\tilde y\equiv\frac{1}{d-2}$ satisfies
    $u(\tilde y)=0$.  Expanding the FP solution around it, $u(y)=a(y-\tilde y)
    +\ot\left((y-\tilde y)^2\right)$,
    and imposing the eigenoperators to be analytic at $\tilde y$,
    $g(y)\sim(y-\tilde y)^n$, the condition for eigenvalues
    equivalent to Eq.\ \ref{LargeN_value} is
    \be\label{Pol_eigen}
      \lambda_n=2(1-\tilde ya)-n(d-2+2\tilde ya).
    \ee

    The GFP corresponds to $u=a=0$, with
    eigenoperators
    \be
      g(y)=(y-\tilde y)^n,\ \ \ n=\frac{2-\lambda}{d-2}.
    \ee

    The HFP corresponds to $h(z)=0$, which gives $a=\frac{(d-2)(4-d)}{2}$ and
    $\lambda=d-2-2n$.

    There is a line of FPs for $d=2+2/n$,
    \be
      y(u)=\frac{C}{u(2-u)}\left(\frac{u}{2-u}\right)^{\frac{d}{2}}+f(u),
    \ee
    which starts at $C=+\infty$ (GFP).
    It ends, for $n$ odd, at $C=0$ (but without reaching the HFP) and, if $n$
    is even,
    at $C=C_0\equiv\frac{\pi d}{\sin\frac{\pi}{n}}$.
    Precisely $C=C_0$, for $n$ even, is the
    only value where the FP is analytical at $u=2$ (it is Taylor-expandable
    around $y=\frac{1}{d+2}$, with $u(\frac{1}{d+2})=2$).
    This allows
    $u=\ot(\frac{1}{y})$ in the vicinity of $y=0$, like
    in the Wegner-Houghton case.
    These non-analyticities associated to the BMB phenomenon seem to present,
    therefore, some kind of universality.

\end{document}